\documentclass[aps,prl, twocolumn, showpacs, superscriptaddress]{revtex4-1}
\usepackage[utf8]{inputenc}
\usepackage{graphicx}
\usepackage{pstricks}
\usepackage{array}
\usepackage{natbib}
\usepackage{amsmath}
\usepackage{pifont}
\usepackage{dsfont}
\usepackage{multirow}
\usepackage{amssymb}
\usepackage{wasysym}

\def\be{\begin{equation}}
\def\ee{\end{equation}}

\begin{document}

\title{Stopping electrons with radio-frequency pulses in the quantum Hall regime}

\newcommand{\spsmsA}{Univ. Grenoble Alpes, INAC-SPSMS, F-38000 Grenoble, France}
\newcommand{\spsmsB}{CEA, INAC-SPSMS, F-38000 Grenoble, France}
\author{Benoit Gaury}
\affiliation{\spsmsA}
\affiliation{\spsmsB}
\author{Joseph Weston}
\affiliation{\spsmsA}
\affiliation{\spsmsB}
\author{Xavier Waintal}
\affiliation{\spsmsA}
\affiliation{\spsmsB}
\date{\today}

\begin{abstract}
Most functionalities of modern electronic circuits rely on the possibility to modify the path
followed by the electrons using, e.g. field effect transistors. Here we discuss the interplay
between the modification of this path and the quantum dynamics of the electronic flow. Specifically, we study 
the propagation of charge pulses through the edge states of a two-dimensional electron gas in the quantum
Hall regime.  By sending radio-frequency (RF) excitations on a top gate capacitively coupled to the electron gas, we manipulate these edge state dynamically. We find that a fast RF change of the gate voltage
can {\it stop} the propagation of the charge pulse inside the sample. This
effect is intimately linked to the vanishing velocity of bulk states in the quantum Hall
regime and the peculiar connection between momentum and transverse confinement of Landau levels. 
Our findings suggest new possibilities for stopping, releasing and switching the trajectory of charge pulses in quantum Hall systems.
\end{abstract}

\maketitle

Electronic states in the quantum Hall regime --- obtained for instance by
applying a strong magnetic field to a two-dimensional heterostructure --- are
very peculiar; with a vanishing velocity in the bulk of the system, they only
propagate (in a chiral way) on the edges of the sample.  Following its initial
discovery some thirty years ago~\cite{Klitzing1980}, the
quantum Hall effect is now used for the metrological measurements of the
quantum of conductance $e^2/h$~\cite{Hartland1991, Janssen2013} as well as a
model system for mesoscopic physics, e.g.  electronic interferometers
~\cite{MachZender_Heiblum, Edge_states_coherence_QHE,Haack}. The corresponding
transport properties can be understood quantitatively in a very simple and
elegant way using the Landauer-B\"uttiker scattering theory and the associated
concept of one-dimensional chiral edge states~\cite{Buttiker1988}.  These edge
states can take place on the actual edges of the sample --- the mesa of the two
dimensional electron gas --- or can be defined by electrostatic gates put on top
of the device. The field effect obtained by applying voltages on these gates is
in turn very peculiar, not only does it allow one to close or open conducting
paths (as in conventional field effect transistors) but it also modifies the
actual paths taken by the electrons or even partitions the edge states into the
superposition of two paths~\cite{MachZender_Heiblum,
Edge_states_coherence_QHE}.

Progress in RF quantum transport are made at an increasing
rate~\cite{McEuen2008,Glattli2013} so that single electron sources have
now moved from theory to the lab~\cite{Single_e_source, Glattli2013}. These
newly available charge sources open a wealth of new possibilities for quantum
electronics.  In this letter,
we discuss the dynamical manipulation of {\it the path taken
by the electron} using fast RF modification of gate voltages. We send charge pulses from 
an Ohmic contact into the system. We find that
these charge pulses can be dynamically manipulated by means of the gates
voltages; they can be stopped, stored and their trajectories switched
dynamically. 

\begin{figure}
    \includegraphics[width=0.47\textwidth]{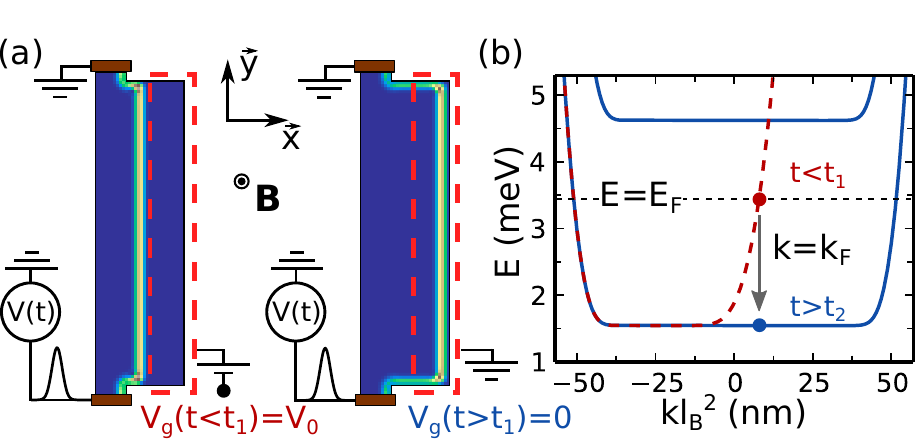}
    \caption{\label{bands} (a) Color maps of $d\rho(x,y)/dV$ of the system
    indicating the position of the edge states at the Fermi level. A gate
    voltage $V_g$ is applied to the electrostatic gate (red dashed rectangle)
    and allows one to shift the position of the edge states: $V_g=V_0$ (left),
    $V_g=0$ (right). (b) Band structure of the system with polarized gate
    ($V_g=V_0$: dashed red) and with grounded gate ($V_g=0$: blue line). The
    times $t_1$ and $t_2$ refer to the
    stopping protocol described in Fig.~\ref{pulse_stop}}
\end{figure}
{\it Mechanism for stopping single electron pulses.} We  start with defining our
``stopping'' protocol and the associated physical mechanism.
Fig.~\ref{bands}(a,b) shows the first (simulated) sample that we consider. A
two-dimensional electron gas (2DEG) under high magnetic field connected to two
Ohmic contacts. We work in a regime where only the lowest Landau levels (LLL)
contribute to the transport properties of the sample. It imposes that all
variations of voltages are slow compared to the cyclotron frequency. The upper
contact is grounded while the lower one is used to send voltage pulses through
the system. A side gate, capacitively coupled to the right-hand side of the
system (dashed line) allows one to modify the  propagating edge states. When the
gate voltage $V_g=V_0$ the current propagates through the middle of the sample
[Fig.~\ref{bands}(a) left] while when the gate is grounded, the current
propagates on the right edge of the sample [Fig.~\ref{bands}(a) right].
Fig.~\ref{bands}(a,b) are not simple schematics of the edge states but
correspond to the extra electronic density $d\rho(x,y)/dV$ that appears
in the 2DEG upon imposing a DC bias voltage $V$ at the lower contact.

\begin{figure*}[t]
    \includegraphics[width=\textwidth]{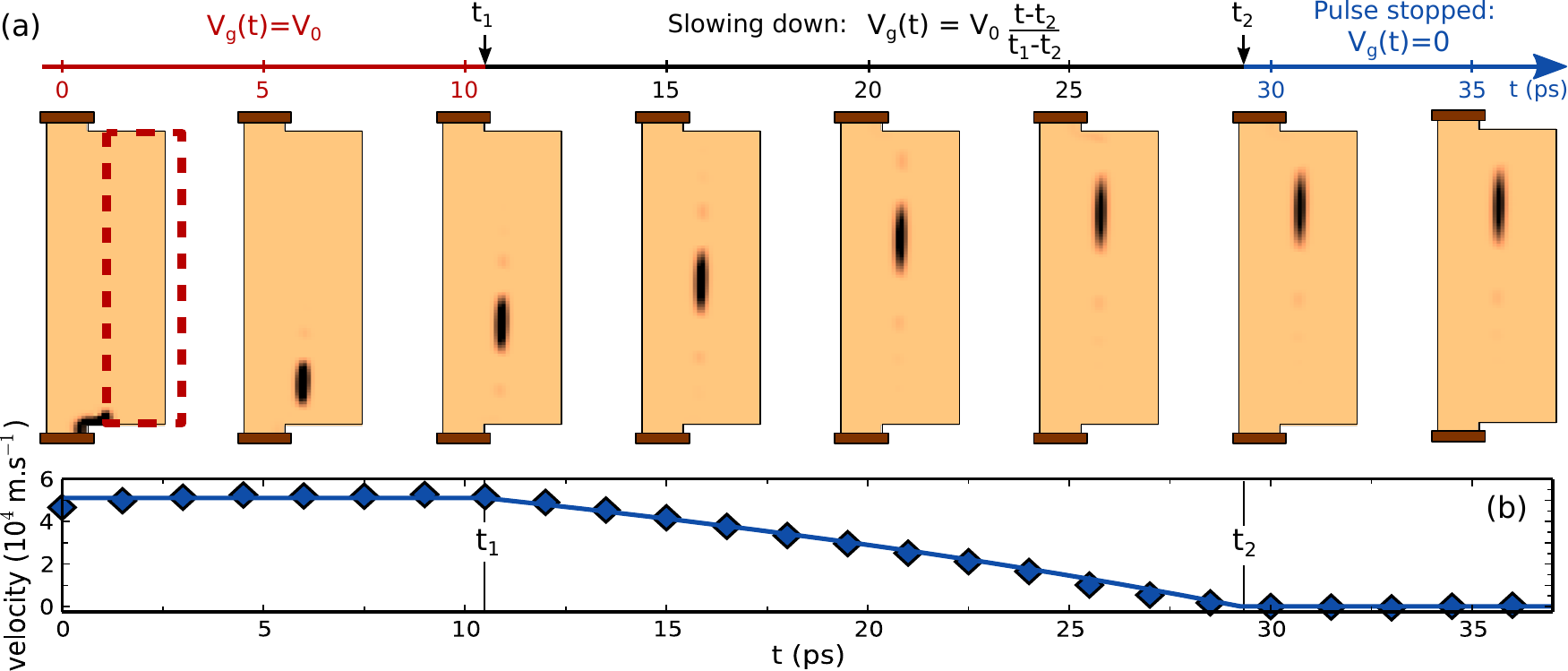}
    \caption{\label{pulse_stop} (a) Color map of the charge density at various
    times during the ``stopping'' protocol. The gate is polarized for $t<t_1$,
    and slowly grounded between $t_1$ and $t_2$. At $t_2$ the pulse is stopped.
    (b) Velocity  $v(t$) of the pulse as a function of time. Diamonds correspond
    to numerical data, the full line to the analytical result.  }
\end{figure*}
The upper part of Fig.~\ref{pulse_stop}(a) shows our ``stopping'' protocol.  
At time $t=0$ we send a voltage pulse $V(t)$ through
the lower contact in presence of a gate voltage $V_g=V_0$ [Fig.~\ref{bands}(a)].
We wait until the pulse has propagated up to
(roughly) one third of the sample and at time $t_1$ we start decreasing the
gate voltage $V_g$. At time $t_2$, $V_g=0$ and the gate is grounded
[Fig.~\ref{bands}(b)]. The snapshots in Fig.~\ref{pulse_stop}(a) show that 
 this protocol actually {\it stops} the propagation of the pulse which stays frozen in
the system for $t>t_2$.

The mechanism behind this behavior can be easily understood from an analysis of
the eigenstates of the system. We model our system with the following
Hamiltonian,
\be
\label{eq:H}
      \mathrm{\hat{\textbf{H}}} =
      \frac{(\vec P -e\vec A)^2}{2m^*} +V(\vec r,t)
\ee
where $\vec P=-i\hbar\vec\nabla$, $\vec A = B x \vec y$ in the Landau gauge,
$B$ is the magnetic field, $m^*$ is the effective mass of the system and the
time-dependent potential $V(\vec r,t)$ contains contributions from the mesa
boundary, the voltage pulse at the Ohmic contact and the electric field due to
the side gate. In the absence of RF pulses, and assuming that our system is
invariant by translation along the y-direction (which it is except close to the
contacts but this is irrelevant), the LLL [that diagonalize Eq.~(\ref{eq:H})]
are localized along the $x$-direction and the plane waves along the $y$-direction
read,
\be
\Psi_{k}(x,y) =  e^{-(x - kl_B^2)^2 / 4l_B^2} \ e^{iky},
\ee
where the magnetic length is defined by $B l^2_B= \hbar/e$. In the absence of
confining potential, the LLL are degenerate with an energy $E(k)=E_0$.
Consequently they are dispersionless with vanishing velocity as $v_k=(1/\hbar)
\partial E/\partial k$. The presence of a confining potential $V(x)$ breaks
this degeneracy. Assuming (for the sake of the argument, our results stand
without this assumption) that $V(x)$ is smooth on the scale of $l_B$, then the
LLL remain eigenstates of the Hamiltonian in presence of the confining
potential and their energy is simply raised by the value of $V(x)$ at the
center of the state, $E(k)=E_0 + V(k l_B^2)$. The corresponding LLL are
propagating on the edges. Fig.~\ref{bands}(b) shows the, numerically calculated,
dispersion relations for $V_g=V_0$ (dashed red) and $V_g=0$ (blue). 

\begin{figure*}[t]
    \includegraphics[width=\textwidth]{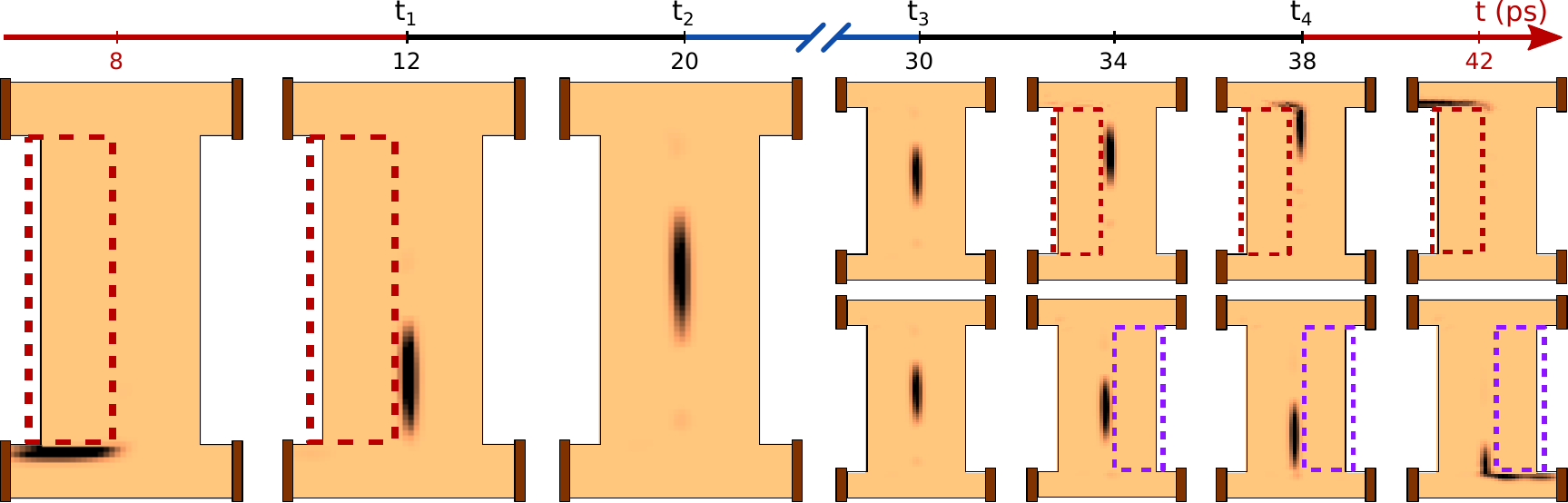}
    \caption{\label{stop-restart} Charge density color map for the ``stop and
    release" protocol.  The two gates on each side of the system (red/blue dashed
    rectangles) control the edge states (hence the direction of propagation of
    the pulse). The left gate is polarized for $t \le t_1$ and grounded for $t
    \ge t_2$.  At $t_2$, the pulse is frozen. At $t=t_3$ one of the two gates
    is polarized again, which releases the pulse.  Top: the left gate is
    polarized, the pulse follows its original edge state and is collected in
    the top left electrode. Bottom: the right gate is polarized, the pulse
    follows the right hand side edge state and is collected  in the bottom
    right electrode.}
\end{figure*}

Let us now go back to the ``stopping'' protocol.  After we have sent the voltage pulse
($0<t<t_1$), the system is in a superposition of LLL with energies close to the
Fermi energy $E_F$ (we use $V(t)\ll E_F$): $\Psi(t)=\sum_k a_k  \Psi_{k}
e^{-iE(k) t}$. At $t>t_1$, we start changing the gate voltage $V_g$. 
Although $V(x,t)$ now depends on time, we should bear in mind that {\it the
system remains invariant by translation along the $y$-direction} at all times. 
As a result
the momentum $k$ is a good quantum number and the linear superposition of LLL
is unmodified. The dispersion relation is now time-dependent with
$E(k,t)=E_0 + V(kl_B^2,t)$ and the wave function reads,
$\Psi(t)=\sum_k a_k  \Psi_{k} e^{-i\int_0^t du E(k,u)}$. In other words, the
energy decreases at fixed momentum $k$, as indicated by the arrow in
Fig.~\ref{bands}(b).  In particular the velocity of the pulse
\be \label{eq:vt}
v(t)=\frac{1}{\hbar} \left.\frac{\partial E(k,t)}{\partial k}\right|_{k_F}
\ee
decreases until it vanishes at $t=t_2$ where the pulse stops. This argument
does not depend on the speed at which the gate voltage is varied as long as it
is fast enough for the pulse not to escape the gated region before the velocity
vanishes. The quantum Hall effect therefore gives us a way to modify
the dispersion relation dynamically and trap particles in a region of vanishing velocity.

{\it Numerical calculations.} We perform direct
numerical simulations of our RF protocol in order to check the above argument.
Equation (\ref{eq:H}) is discretized on a lattice according to usual
prescriptions~\cite{knit} with standard parameters for $GaAS/AlGaAs$
heterostructures. We consider a 2DEG of density $n_s=10^{11}cm^{-2}$, 
corresponding to a Fermi energy $E_F=3.47meV$ or equivalently to a Fermi wave
length $\lambda_F=79nm$. A magnetic field $B=1.8T$ is applied to the system
yielding a magnetic length $l_B =19nm$ and a cyclotron frequency
$\hbar\omega_c=3.1meV$ ($\omega_c=eB/m^*$). We used a realistic confining
potential for the gate that corresponds to a drift velocity
$v=5 \ 10^4 m.s^{-1}$ but we did not actually solve the associated
electrostatics. DC calculations are performed with
Kwant~\cite{Kwant_preparation}. RF simulations are performed with
T-Kwant~\cite{Twave_formalism, fabryperot}.

In Fig.~\ref{pulse_stop}(a), 
a Gaussian pulse
$V(t)=V_p \exp(-4\log(2)t^2/\tau_p^2)$ of duration $\tau_P=2ps$
and amplitude $V_P=0.4mV$ is sent through the system. Fig.~\ref{pulse_stop}(a) actually shows the
difference between two simulations performed with and without the voltage
pulse .
Indeed, upon decreasing $V_g$, the system relaxes to a new equilibrium (with
electrons entering the system in order to fill the formerly forbidden region).
We discuss this aspect briefly toward the end of this letter. As
expected, we find that the pulse is indeed stopped for $t>t_2$. More importantly,
Fig.~\ref{pulse_stop}(b) shows a quantitative agreement between the numerics and
the analysis made above. The symbols show the velocity of the pulse as
measured from the time-dependent numerics (by looking at the time evolution of
the center of mass of the electronic density carried by the pulse) while the
line corresponds to Eq.~(\ref{eq:vt}).

{\it ``Stop and release'' protocol.} Now that we have established the 
mechanism for stopping the pulse, we proceed with a slightly different sample
with 4 terminals and an additional top gate, see Fig.~\ref{stop-restart}. 
The first part of the protocol of Fig.~\ref{stop-restart} is the
same as previously. We sent a pulse at $t=0$ (now the pulse is sent from the
lower left contact and the left gate is polarized) and stop it by gradually
grounding the left gate between $t_1$ and $t_2$. For $t_2<t<t_3$ the voltage
pulse is stuck in the middle of the sample. After waiting for some time, until
$t_3$, we do one of two things. Either we increase again the voltage of the
left gate (upper panels) in order to restart the pulse, or we increase the
voltage of the other (right) gate (lower panels) which also restarts the pulse
but in a different direction. From a theoretical point of view, both cases are
very similar and are essentially the counter-part of the stopping
protocol (and can be analyzed accordingly). However, in practice they
illustrate the versatility of what could be accomplished with this dynamical
modification of the paths of the electrons. This RF protocol allows one to stop
a charge pulse, then store it for a while in a region with vanishing velocity, and
finally release it in a direction of our choice. 

{\it Probing ``stopped and released'' pulses with a Mach-Zehnder
interferometer.} We end this letter with a last set of simulations that allows
\begin{figure}[h]
    \includegraphics[width=0.47\textwidth]{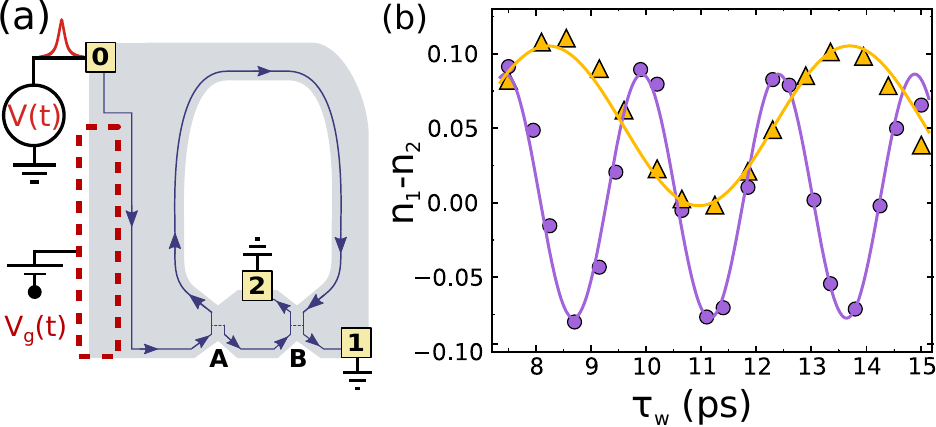}
    \caption{\label{stpmz} (a) Schematic of a Mach-Zehnder interferometer with a
    ``stop and release" gate. The blue line shows the two paths of the
    interferometer. (b) Difference $n_1-n_2$ between the transmitted charges
    into contacts 1 and 2 as a function of the waiting time of the pulse
    $\tau_w$ for $E_F=3.47meV$ (purple circles) and $E_F=2.5meV$ (yellow
    triangles). Lines correspond to the fit $n_1-n_2=a_1 + a_2
    \sin^2([(E_F-E_0)/2] \tau_w)$.}
\end{figure}
us to further analyze the nature of the ``stop and release'' protocol. In the
sample sketched in Fig.~\ref{stpmz}(a), we send a voltage pulse, stop it with a
gate (as previously), wait for some time $\tau_w$, and release the pulse
(again, as previously).  However, instead of directly collecting the current in
the electrode, it is sent through an electronic Mach-Zehnder interferometer
obtained with two half-transmitting quantum point contacts A and B. We refer
to~\cite{MachZender_Heiblum, Edge_states_coherence_QHE} (\cite{knit,fabryperot})
for experimental (theoretical) details about the Mach-Zehnder geometry.
Fig.~\ref{stpmz}(b) shows the difference between the total number of electrons
collected at electrodes 1 ($n_1$) and 2 ($n_2$)
as a function of the
waiting time $\tau_w$.  The result is at first sight rather intriguing,
$n_1-n_2$ {\it oscillates} with $\tau_w$ as $\cos( (E_F-E_0)\tau_w )$.  To
understand this behavior, one needs to remember that a voltage pulse is not
simply a localized charge pulse propagating in vacuum, indeed a delocalized
plane wave (LLL) $\Psi \propto e^{iky-iEt}$ already exists before the pulse is
sent.  As one raises the bias voltage $V(t)$, the part of the wave at higher
voltage starts accumulating an extra phase $\phi (t)=\int^t du\ eV(u)/\hbar$.
Noting that $\phi(\infty)=2\pi\bar n$ ($\bar n$: number of injected particles) 
and supposing the voltage drop to be
concentrated around $y=0$, the wave function just after the pulse takes
the form $\Psi \propto e^{iky-i2\pi\bar n \theta(-y)}$ where $\theta(y)$ is the
Heaviside function.  In other words a voltage pule generates a kink in the
phase of the wave function.  This kink, or phase domain wall, carries charges
and propagates ballistically through the sample.  The $2\pi\bar n$ phase
difference 
between the front and the rear of the pulse causes oscillations of
$n_1-n_2$ with $\bar n$ owing to the ``dynamical control of interference
pattern"~\cite{fabryperot}.  We now come back to our ``stop and release"
protocol (ignoring the presence of the voltage pulse). We suppose that the part
of the edge state which is affected by the gate corresponds to $y\in [0,L ] $
(using curved coordinates that follow the edge state). Before $t_1$, we have a
plane wave $\Psi \propto e^{iky-iEt}$. After $t_2$, the
inner part for $y\in [0,L ] $ oscillates as $e^{iky-iE_0t}$ while the rest of
the wave, unaffected by the gate, still oscillates as $e^{iky-iEt}$. Therefore,
after the waiting time $\tau_w$, a phase difference $2\pi\bar
n_w=(E-E_0)\tau_w$ has been accumulated between the inner part and the outer
one. When one releases the pulse again at time $t_3$, the wave function reads
$\Psi \propto e^{iky+i2\pi\bar n_w \theta(y)\theta(L-y)}$. In other words, the
``stop and release'' procedure is equivalent to introducing {\it two} voltage pulses in
series separated by a distance $L$, one effective pulse of $\bar n_w$ electrons
followed by a counter-pulse of $-\bar n_w$ electrons. The oscillation shown in
Fig.~\ref{stpmz}(b) simply follows from the ``dynamical control of interference
pattern" of~\cite{fabryperot} applied to this series of two pulses.

{\it Qualitative discussion of charge relaxation in the system}. 
Let us briefly discuss what happens in the ``stopping'' protocol
when one does not send any voltage pulse in the system.  We  suppose,
for the sake of the argument, that the gate is grounded very abruptly
($t_2=t_1$).  Just after $t_2$ the former edge state is frozen as discussed
extensively above.  On the other hand, a new one (which was at very high energy
before $t_2$) now appears on the edge of the mesa. This edge state is initially
empty and gets gradually filled as electrons pour in from the electrode. In 
our
non-interacting model, only the propagating modes get filled in,
leaving an empty puddle in the region of the 2DEG where the velocity vanishes (most
of the area under the gate). This is of course unphysical as it raises the
electrostatic energy of the system.  
As the new edge state is filled,
the corresponding charges create a local electric field; the neighboring
edge states become dispersive, and start to get filled as well. This process
continues until all the LLL below the gate are filled and the system has
relaxed to its equilibrium. This relaxation process should be very slow as the whole
area underneath the gate needs to be filled while the electrons can only be
poured in through one-dimensional edge states. A proper treatment of this
physics would require solving the Poisson equation self-consistently with
quantum mechanics. It would  allow one to describe the charge relaxation using the
compressible and incompressible regions discussed in~\cite{Chlovskii1992}.
We expect however that the (current carrying) compressible stripes behave
essentially in the same way as the edge states of the non-interacting theory
used in this letter. Simulations of these phenomena will be the subject
of future work. In any case, performing the difference between two simulations (with and without charge pulse) 
allows us to disentangle the pulse physics (of interest here) from the charge relaxation (poorly described by our model). A similar protocol should be followed experimentally.

{\it Conclusion.} RF quantum electronics is a very young emerging field. Here,
we have presented a few possibilities offered in the quantum Hall regime. Even
in the simplest situation, one predicts intriguing, often counter intuitive,
results~\cite{fabryperot}. We note that the practical implementation of the proposals presented in this letter
imply delicate experiments where one injects high frequency pulses in a dilution
fridge setup. The measurement scheme however should not be too difficult as, by
periodically repeating the pulse sequences, measuring the number of electrons
received in one electrode amounts to measuring DC currents.

{\it Acknowledements.} This work is funded by the ERC consolidator grant
MesoQMC. We thank C. Bauerle, C. Glattli, L. Glazmann, F. Portier and P. Roche for useful discussions.

\bibliographystyle{apsrev}
\bibliography{references}

\end{document}